\documentclass[twocolumn,prl,amsmath,amssymb]{revtex4-1}
\usepackage{geometry}
\usepackage{color}
\usepackage{graphicx}
\begin{document}


\title{Electronic Structure of Ce-Doped and -Undoped Nd$_2$CuO$_4$ Superconducting Thin Films Studied by Hard X-ray Photoemission and Soft X-ray Absorption Spectroscopy}

\author{M. Horio$^1$, Y. Krockenberger$^2$, K. Yamamoto$^3$, Y. Yokoyama$^3$, K. Takubo$^3$, Y. Hirata$^3$, S. Sakamoto$^1$, K. Koshiishi$^1$, A. Yasui$^4$, E. Ikenaga$^4$, S. Shin$^3$, H. Yamamoto$^2$, H. Wadati$^3$, and A. Fujimori$^1$}

\affiliation{$^1$Department of Physics, University of Tokyo, Bunkyo-ku, Tokyo 113-0033, Japan}
\affiliation{$^2$NTT Basic Research Laboratories, NTT Corporation, Atsugi, Kanagawa 243-0198, Japan}
\affiliation{$^3$Institute for Solid State Physics, University of Tokyo, Kashiwa, Chiba 277-8561, Japan}
\affiliation{$^4$Japan Synchrotron Radiation Research Institute, Sayo, Hyogo 679-5198, Japan}

\date{\today}

\begin{abstract}
In order to realize superconductivity in cuprates with the T'-type structure, not only chemical substitution (Ce doping) but also post-growth reduction annealing is necessary. In the case of thin films, however, well-designed reduction annealing alone without Ce doping can induce superconductivity in the T'-type cuprates. In order to unveil the origin of superconductivity in the Ce-undoped T'-type cuprates, we have performed bulk-sensitive hard x-ray photoemission and soft x-ray absorption spectroscopies on superconducting and non-superconducting Nd$_{2-x}$Ce$_x$CuO$_4$ ($x = 0, 0.15$, and 0.19) thin films. By post-growth annealing, core-level spectra exhibited dramatic changes, which we attributed to the enhancement of core-hole screening in the CuO$_2$ plane and the shift of chemical potential along with changes in the band filling. The result suggests that the superconducting Nd$_2$CuO$_4$ film is doped with electrons despite the absence of the Ce substitution.
\end{abstract}


\maketitle

High-temperature superconductivity in cuprates is realized by doping hole or electron carriers into the parent material which has been widely considered to be an antiferromagnetic (AFM) Mott insulator. $Ln_2$CuO$_4$ ($Ln$: rare earth) with the T'-type structure, where Cu takes the square-planar coordination, can be doped with electrons by substituting Ce$^{4+}$ for $Ln^{3+}$. Generally, electron doping by Ce substitutions alone cannot induce superconductivity in bulk crystals of the T'-type cuprates, and an additional procedure of post-growth annealing in a reducing atmosphere is required \cite{Tokura1989}. Three distinct microscopic scenarios have been experimentally proposed for the role of the reduction annealing \cite{Armitage2010}: i) removal of impurity apical oxygen atoms \cite{Radaelli1994,Schultz1996} ii) creation of vacancies at the regular oxygen sites \cite{Riou2004,Richard2004} iii) filling of Cu vacancies \cite{Kang2007}. While the exact microscopic effect remains elusive, reduction annealing dramatically suppresses the AFM order \cite{Richard2007,Horio2016} and reduces quasi-particle scattering \cite{Xu1996}, inducing superconductivity in the Ce-doped samples. 

Furthermore, it has been demonstrated that thin films of T'-type cuprates can exhibit superconductivity without any Ce doping when properly annealed \cite{Tsukada2005a,Tsukada2005b,Matsumoto2009b,Krockenberger2013}. The observed decrease of the $c$-axis parameter by annealing  \cite{Tsukada2005b,Matsumoto2009b,Krockenberger2013} rather than its increase \cite{Kang2007} should be attributed to the removal of apical oxygen atoms in these samples \cite{Radaelli1994,Schultz1996}, as the removal of apical oxygen reduces the repulsion between the CuO$_2$ planes. The large surface-to-volume ratio of thin films helps the thorough removal of impurity oxygen atoms. The observation has cast doubt on the fundamental assumption that the parent compound of the cuprate superconductor is an AFM Mott insulator. Theoretical studies using the local density approximation plus dynamical mean field theory (LDA+DMFT) have indeed predicted that the parent compound of the T'-type cuprates is not a Mott insulator but a Slater insulator in the sense that the AFM order is necessary to open the insulating band gap \cite{Das2009,Weber2010a,Weber2010b}. When discussing the electronic structure of parent compounds and the phase diagram, however, the possibility should not be overlooked that oxygen reduction affects the carrier concentration. In fact, annealing-induced changes in the N\'{e}el temperature \cite{Mang2004a} and the optical conductivity \cite{Arima1993} for bulk Nd$_{2-x}$Ce$_x$CuO$_4$ crystals have been interpreted as due to the doping 0.03--0.05 electrons/f.u. Recent angle-resolved photoemission spectroscopy (ARPES) studies on Ce-doped bulk single crystals \cite{Horio2016,Song2017} and that on insulating Ce-undoped thin films \cite{Wei2016} have also shown that the electron concentrations of annealed samples estimated from the Fermi surface area are larger than that expected from the Ce concentrations. To understand the cuprate phase diagram and the electronic states of the parent compounds, it is important to unveil the electron concentration of the superconducting (SC) and non-superconducting (non-SC) Ce-undoped T'-type cuprates. 

For that purpose, systematic studies of thin film samples with various Ce concentrations and reduction/oxidization treatments are necessary. However, ARPES measurements on such thin films are hampered by the lack of an {\it in-situ} ARPES measurement system combined with a molecular beam epitaxy (MBE) apparatus for the T'-type cuprates, which makes ARPES-quality surfaces available. This is because ARPES is a surface-sensitive technique. In this Letter, we report on measurements of bulk-sensitive hard x-ray photoemission spectroscopy (HAXPES) and soft x-ray absorption spectroscopy (XAS) of SC and non-SC Nd$_{2-x}$Ce$_x$CuO$_4$ thin films annealed or oxidized under various atmosphere. By HAXPES, one can measure core-level shifts and hence the chemical-potential shift, which directly probes the doped electron concentration. By annealing Nd$_2$CuO$_4$, we observed dramatic changes in the core-level spectra, which can be explained by the strong modification of core-hole screening in the CuO$_2$ planes and by the chemical-potential shift caused by electron doping. The present results indicate the possibility that the SC Ce-undoped T'-type cuprates are doped with a significant amount of electrons.

\begin{figure}
\begin{center}
\includegraphics[width=85mm]{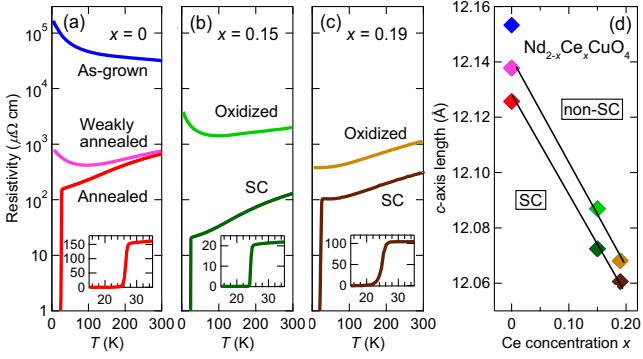}
\end{center}
\caption{Physical properties of Nd$_{2-x}$Ce$_x$CuO$_4$ thin films. (a)--(c) Resistivity versus temperature for the $x = 0$, 0.15, and 0.19 films, respectively. The inset shows a magnified plot near the SC transition for each composition. (d) $c$-axis lengths plotted against Ce concentration $x$. The markers are color-coded according to (a)--(c). Two solid lines trace the $c$-axis lengths of SC and non-SC films whose difference may originate from the different amount of apical oxygen atoms.}
\label{NCCO_rhoT}
\end{figure}

Nd$_{2-x}$Ce$_x$CuO$_4$ ($x = 0, 0.15$, and 0.19) thin films with the thicknesses of 200 nm, 100 nm, and 100 nm, respectively, were grown on SrTiO$_3$ (001) substrates by MBE. For Nd$_2$CuO$_4$, we prepared three kinds of films: as-grown, weakly-annealed, and annealed films, among which only the annealed film showed superconductivity with $T_\mathrm{c} = 25.0$ K. The $T_\mathrm{c}$ was defined as the temperature where the resistivity drops to zero. Ce-doped films showed superconductivity without \textit{ex-situ} annealing in a tubular furnace. The $T_\mathrm{c}$'s were 24.0 K and 21.5 K for $x = 0.15$ and $x = 0.19$, respectively. Oxidized non-SC films were also prepared for both the compositions. Conditions of annealing and oxidization are described in Supplementary Information. The resistivity curves and the $c$-axis lengths of all the films are plotted in Fig.~\ref{NCCO_rhoT}. The difference in the $c$-axis length between the SC and non-SC films in Fig.~\ref{NCCO_rhoT}(d) may largely originate from the difference in the amount of apical oxygen atoms \cite{Tsukada2005b,Matsumoto2009b,Krockenberger2013}. HAXPES measurements were performed at beamline 47XU of SPring-8 at $T = 300$ K with $h\nu = 7.94$ keV photons. The total energy resolution was determined from the Fermi edge of Au to be 0.3 eV. XAS measurements were performed in the total electron yield mode at beamline 07LSU of SPring-8 at $T = 300$ K under the pressure better than $5 \times 10^{-9}$ Torr. Two kinds of linearly polarized light, with polar angle $\theta = 90^\circ$  (${\bf E} \perp c$) and $\theta = 30^\circ$, were used for the measurements \cite{Sup1}.

\begin{figure}
\begin{center}
\includegraphics[width=85mm]{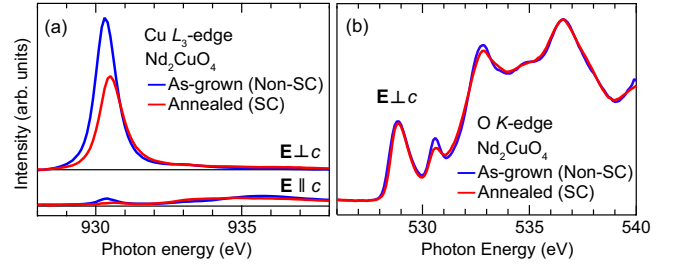}
\end{center}
\caption{XAS spectra of Nd$_2$CuO$_4$ thin films. (a) Cu $L_3$-edge XAS spectra for ${\bf E} \perp c$ (top) and ${\bf E} \parallel c$ (bottom) polarizations. The spectra have been normalized to the intensity of the Nd $M_{4,5}$ XAS peak at 978 eV. (b) O $K$-edge XAS spectra for ${\bf E} \perp c$ and normalized to the intensity at 532--540 eV.}
\label{XAS}
\end{figure}

Figure~\ref{XAS}(a) shows Cu $L_3$-edge XAS spectra for ${\bf E} \parallel c$ and ${\bf E} \perp c$ for as-grown and annealed Nd$_2$CuO$_4$ films. The spectra for ${\bf E} \parallel c$ were obtained by subtracting the contribution of  ${\bf E} \perp c$ from the spectra measured with $\theta = 30^\circ$ polarization. The absorption intensity is proportional to the unoccupied density of states (DOS) of the specific element multiplied by transition matrix elements \cite{deGroot1990}, which depends on the initial- and final-state orbital symmetry as well as incident light polarization. Thus, XAS provides information about the element- and orbital-specific unoccupied DOS. With polarization ${\bf E} \perp c$, matrix element of the transition into the 3$d_{x^2-y^2}$ orbital is three times larger than that into the 3$d_{3z^2-r^2}$ orbital, while the polarization ${\bf E} \parallel c$ allows only the transition into the 3$d_{3z^2-r^2}$ orbital (See Supplementary Information). The XAS spectra for ${\bf E} \perp c$ show an intense peak at $\sim 930$ eV and its intensity decreases with annealing by 30 \%, suggesting the reduction of unoccupied DOS near $E_\mathrm{F}$ as reported for Ce-doped samples \cite{Flipse1990,Pellegrin1993,Fink1994,Steeneken2003}. In contrast, spectra for ${\bf E} \parallel c$ remains negligibly weak. This leads to a small ratio of the 3$d_{3z^2-r^2}$ weight to the 3$d_{x^2-y^2}$ weight (4 \% and 2 \% for as-grown and annealed Nd$_2$CuO$_4$, respectively) in agreement with previous reports \cite{Pellegrin1993,Fink1994}. Therefore, the 3$d_{3z^2-r^2}$ orbitals are almost completely filled regardless of whether annealed or not, and additional electrons go to the 3$d_{x^2-y^2}$ orbitals by annealing. On the other hand, the pre-edge peak ($\sim$ 529 eV) in O $K$-edge XAS measured with ${\bf E} \perp c$ [Fig.~\ref{XAS}(b)], which represents the transition from O 1$s$ to in-plane O 2$p$ orbitals hybridized with the upper Hubbard band, shows only a slight change (5 \%) in contrast to Cu $L_{2,3}$-edge XAS for ${\bf E} \perp c$. This agrees with the changes observed by Ce doping in the previous studies and suggests that the orbital character of the upper Hubbard band is dominated by 3$d_{x^2-y^2}$ \cite{Flipse1990,Krol1990,Pellegrin1993,Fink1994}.

\begin{figure*}[ht!]
\begin{center}
\includegraphics[width=180mm]{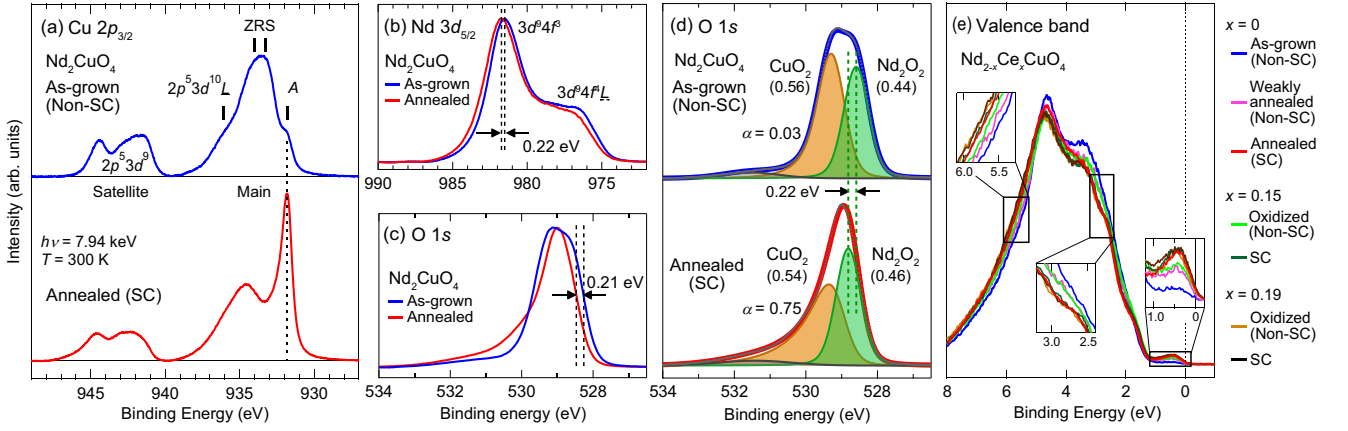}
\end{center}
\caption{HAXPES spectra of Nd$_{2-x}$Ce$_x$CuO$_4$ thin films. (a) Cu 2$p_{3/2}$ core-level spectra of the as-grown non-SC (top) and annealed SC Nd$_2$CuO$_4$ thin films (bottom). One observe final states where the core holes are unscreened (2$p^53d^9$), screened by an electron transferred from neighboring oxygen atoms (2$p^53d^{10} \underline{L}$), screened by an electron transferred from neighboring CuO$_4$ plaquettes thereby creating a Zhang-Rice singlet (ZRS), and screened by conduction electrons (A) (b),(c) Nd 3$d_{5/2}$ and O $1s$ core-level spectra normalized to the peak height, respectively, for the as-grown and annealed Nd$_2$CuO$_4$ films. (d) O $1s$ spectra of the as-grown (top) and annealed (bottom) Nd$_2$CuO$_4$ films fitted to a superposition of a Voigt function (for O$_\mathrm{Nd_2O_2}$, green), a Mahan line shape $\frac{1}{\mathrm{\Gamma}(\alpha)} \frac{e^{-(E_B-E_0)/\xi}}{\left| (E_B-E_0)/\xi \right| ^{1-\alpha}} \Theta(E_B-E_0)$ convolved with a Voigt function (for O$_\mathrm{CuO_\mathrm{2}}$, orange), and another Voigt function (for O contamination, gray). Values in the parentheses represents the ratio of the peak area. (e) Valence-band spectra of Nd$_{2-x}$Ce$_x$CuO$_4$ thin films.}
\label{Nd_O}
\end{figure*}

The effect of annealing is also remarkable in HAXPES spectra. Figure~\ref{Nd_O}(a) shows the Cu 2$p_{3/2}$ core-level HAXPES spectra of the as-grown and annealed Nd$_2$CuO$_4$ films. Upon photoemission from a core level, valence electrons are attracted to the core-hole site to screen its potential. Different types of core-hole screening result in various final states appearing as fine structures in the core-level photoemission spectra \cite{Veenendaal1993}. As has been discussed for the Cu 2$p$ spectrum of insulating cuprates such as La$_2$CuO$_4$ \cite{Taguchi2005,Maiti2009} and Sr$_2$CuO$_2$Cl$_2$ \cite{Boske_1997,Boske_1998} analyzed using multi-site cluster calculation, the peak at the highest binding energy ($E_\mathrm{B}$ = 940--947 eV) is the satellite due to the 2$p^5$3$d^9$ final state with an unscreened core hole. In the main peak region ($E_\mathrm{B}$ = 930--939 eV), the highest binding energy peak corresponds to the 2$p^53d^{10} \underline{L}$ final state, where an electron is transferred from the neighboring oxygen atoms (local screening) leaving a hole in the oxygen ligand orbital $L$, and the two peaks in the middle to the final state where an electron is transferred from oxygen in the neighboring CuO$_4$ plaquettes thereby creating a Zhang-Rice singlet in the plaquettes (non-local screening). Apart from peak A at the lowest binding energy discussed below, the line shape of the main peak resembles those of the other insulating cuprates with two-dimensional CuO$_2$ planes \cite{Boske_1998}.

Upon annealing, peak A is strongly enhanced and the entire line shape dramatically changes. The binding energies of the charge-transferred final state are determined by the energy levels of the electronic states from which the electron is transferred to screen the core-hole potential. Therefore, the enhancement of the lowest energy peak indicates the development of electronic states closest to the Fermi level ($E_\mathrm{F}$) \cite{Taguchi2005,Panaccione2008} and is attributed to final states where the core hole is screened by conduction electrons. The observed changes in the present Cu 2$p$ spectra therefore suggest that conduction electrons were introduced by annealing, consistent with the occurrence of superconductivity in the annealed Nd$_2$CuO$_4$.


The effect of annealing on the Nd and O core levels of Nd$_2$CuO$_4$ films is shown in Figs.~\ref{Nd_O}(b)--(d). Upon annealing, the Nd 3$d_\mathrm{5/2}$ peak was shifted toward higher binding energy by 0.22 eV [Fig.~\ref{Nd_O}(b)]. As for the O $1s$ peak, a shift of similar amount (0.21 eV) was observed at the edge, but the spectral line shape is also modified: the full width at half maximum decreased from 1.44 eV to 1.16 eV and a long high binding energy tail emerges above $\sim 529.5$ eV [Fig.~\ref{Nd_O}(c)]. Such changes have been overlooked in the doping and annealing dependence of the O $1s$ photoemission spectra of Nd$_{2-x}$Ce$_x$CuO$_4$ in previous studies \cite{Suzuki1990a,Harima2001}. In order to disentangle this complicated spectral deformation, the O $1s$ spectra have been analyzed as follows: The peak region of the O $1s$ spectrum of the as-grown film is rather flat, implying unresolved two peaks with similar intensities. We attribute them to oxygen atoms in the Nd$_2$O$_2$ layers (O$_\mathrm{Nd_2O_2}$) and those in the CuO$_2$ planes (O$_\mathrm{CuO_\mathrm{2}}$), which are contained in Nd$_2$CuO$_4$ with equal amount. We calculated the oxygen binding energies using a Wien2k package, and found that the O$_\mathrm{Nd_2O_2}$ 1$s$ level was located at a lower binding energy than the O$_\mathrm{CuO_\mathrm{2}}$ 1$s$ level \cite{Sup2}. Therefore, the O 1$s$ spectra of the Nd$_2$CuO$_4$ films were fitted to a superposition of a Voigt function (for O$_\mathrm{Nd_2O_2}$) at lower binding energies and Mahan line shape $\frac{1}{\mathrm{\Gamma}(\alpha)} \frac{e^{-(E_B-E_0)/\xi}}{\left| (E_B-E_0)/\xi \right| ^{1-\alpha}} \Theta(E_B-E_0)$ convolved with a Voigt function (for O$_\mathrm{CuO_\mathrm{2}}$) at higher binding energies. The asymmetric Mahan line shape was assumed for O$_\mathrm{CuO_\mathrm{2}}$ considering the core-hole screening by metallic electrons. Another Voigt function was added to the fitting function to reproduce a weak contamination component at $\sim 531$ eV. The fitting yielded O$_\mathrm{Nd_2O_2}$ and O$_\mathrm{CuO_\mathrm{2}}$ peaks with nearly the same area both for the as-grown and annealed films as shown in Fig.~\ref{Nd_O}(d), consistent with the initial assumption that each O 1$s$ spectrum consists of two components arising from O$_\mathrm{CuO_2}$ and O$_\mathrm{Nd_2O_2}$ \cite{Sup3}. The present analysis thus enables us to identify the effect of annealing on the two O 1$s$ core levels separately: The O$_\mathrm{Nd_2O_2}$ peak was shifted by 0.22 eV toward higher binding energy, and the O$_\mathrm{CuO_\mathrm{2}}$ peak became strongly asymmetric (represented by the increase of asymmetry parameter $\alpha$ from 0.03 to 0.75). Since a finite DOS at $E_\mathrm{F}$ leads to a peak asymmetry, the strong asymmetry of the O$_\mathrm{CuO_\mathrm{2}}$ peak in the annealed Nd$_2$CuO$_4$ film is consistent with the dramatic enhancement of the electrical conductivity. Another remarkable point is that annealing shifted the O$_\mathrm{Nd_2O_2}$ 1$s$ peak by almost the same amount as the Nd 3$d_\mathrm{5/2}$ peak without appreciable changes in the line shapes.

The shift of the core-level binding energy is given by \cite{Hufner1995}
\begin{equation}
\Delta E_\mathrm{B} = \Delta \mu - K\Delta \mathrm{Q} + \Delta V_\mathrm{M} - \Delta E_\mathrm{R},
\label{shift}
\end{equation}
where $\Delta \mu$ is the change in the chemical potential, $\Delta Q$ is the change in the number of valence electrons, $K$ is a constant, $\Delta V_\mathrm{M}$ is the change in the Madelung potential, and $\Delta E_\mathrm{R}$ is the change in the extra-atomic screening of the core-hole potential by conduction electrons and/or dielectric polarization of surrounding media. Almost identical shifts observed for the Nd 3$d$ and O$_\mathrm{Nd_2O_2}$ 1$s$ core levels indicate that $\Delta V_\mathrm{M}$ is negligibly small because it would shift the core levels of the O$^{2-}$ anion and the Nd$^{3+}$ cation in different ways. $\Delta E_\mathrm{R}$ cannot be the main origin of the observed shifts, either, because the increase of conduction electrons by annealing would shift the core-level peaks toward lower binding energy, opposite to the experimental observation. Considering that the valences of Nd$^{3+}$ and O$^{2-}$ are fixed ($\Delta Q = 0$), we conclude that the observed shifts in Nd 3$d$ and O$_\mathrm{Nd_2O_2}$ 1$s$ core levels of Nd$_2$CuO$_4$ are largely due to the chemical potential shift $\Delta \mu$. The increase of the core-level binding energies by annealing indicates the increase of $\Delta \mu$ due to the addition of electrons.

\begin{figure}
\begin{center}
\includegraphics[width=85mm]{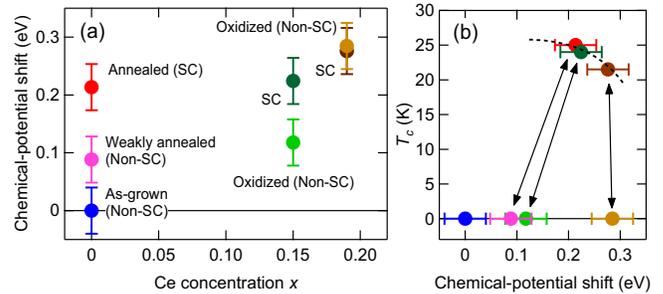}
\end{center}
\caption{Chemical-potential shifts in Nd$_{2-x}$Ce$_x$CuO$_4$ thin films. (a) Chemical-potential shifts $\Delta \mu$ defined as the average of the Nd 3$d$ and O 1$s$ core-level shifts plotted against Ce concentration $x$. (b) $T_\mathrm{c}$'s plotted against $\Delta \mu$. The markers are color-coded according to (a). Thin films with shorter (longer) $c$-axis lengths and hence less (more) apical oxygen atoms are SC (non-SC) [See Fig.~\ref{NCCO_rhoT}(d)]. The arrows connect the films with the same Ce concentration.}
\label{muShift}
\end{figure}

Having identified the chemical-potential shift caused by annealing in Nd$_2$CuO$_4$, we compare the core-level structure of Nd$_2$CuO$_4$ with those of Ce-doped compounds. By Ce substitution, annealing, and oxidization, the Nd and Ce 3$d$ spectra were shifted maintaining their shape. The line-shape changes in the Cu 2$p$ and O 1$s$ spectra were also consistent with the above scenario \cite{Sup4}. The chemical-potential shift $\Delta \mu$ defined as the average shift of the Nd 3$d$ and O$_\mathrm{Nd_2O_2}$ 1$s$ core levels, is plotted in Fig,~\ref{muShift}(a) against Ce concentration $x$. For Nd$_2$CuO$_4$, the chemical potential is shifted upwards with annealing, reaching the level of Ce-doped $x=0.15$ and 0.19 superconductors when sufficiently annealed. The valence-band spectra shown in Fig.~\ref{Nd_O}(e) and Supplementary Information are also almost identical among the three SC films with different Ce concentrations ($x=0$, 0.15, and 0.19), indicating that the electronic structure and band filling are close to each other.

The large electron concentration for the annealed SC Nd$_2$CuO$_4$ film can be explained if oxygen atoms are removed not only from the apical site \cite{Radaelli1994,Schultz1996} but also from the regular sites (in the CuO$_2$ plane and/or the Nd$_2$O$_2$ layer) \cite{Riou2004,Richard2004}, leading to the total oxygen content less than the stoichiometric one. On the other hand, upon oxidization of Nd$_{2-x}$Ce$_x$CuO$_4$ ($x = 0.15$), chemical potential was shifted downwards as can be seen in Fig.~\ref{muShift}(a). This may be because the SC $x=0.15$ film was already oxygen deficient and the vacancies were filled by oxidization. Consequently, the band filling of the T'-type cuprates is not determined by the Ce substitution alone but by the combined effect of Ce substitution and oxygen vacancies and, therefore, the phase diagram for this system should be made as a function of actual band filling, rather than as a function of Ce concentration. The oxidization process also introduced excess apical oxygen atoms, as suggested by the $c$-axis elongation [Fig.~\ref{NCCO_rhoT}(d)], and the $x=0.15$ film turned non-SC.

In Fig.~\ref{muShift}(b), the $T_\mathrm{c}$'s of Nd$_{2-x}$Ce$_x$CuO$_4$ films are thus plotted against chemical-potential shift, which represents the electron concentration. The $T_\mathrm{c}$ values are not solely determined by the electron concentration since $x = 0.19$ films can be either SC or non-SC without a significant change in the electron concentration. The elongation of the $c$-axis lattice parameter by oxidization is also rather small for the $x = 0.19$ film (0.06 \%) compared to that for $x = 0.15$ film (0.12 \%), indicating that only a tiny amount of apical oxygen atoms were incorporated. These results are consistent with previous studies where the amount of oxygen reduction decreases with increasing Ce concentrations \cite{Suzuki1990b,Radaelli1994,Schultz1996}. The higher concentration of Ce$^{4+}$ and the smaller $c$-axis lattice parameter may make the Ce-overdoped samples more robust against oxygen non-stoichiometry induced by reduction annealing and oxidization.
While oxygen vacancy at the regular sites increases the electron concentration, excess oxygen atoms at the apical site immediately destroy superconductivity. The electronic structure of the T'-type cuprates are thus dominated not only by Ce concentrations but also by oxygen non-stoichiometry.

In conclusion, we have performed HAXPES and XAS measurements on Nd$_{2-x}$Ce$_x$CuO$_4$ ($x$ = 0, 0.15, and 0.19) thin films with varying annealing atmosphere, and observed changes in the band-filling level among them. The electronic structure of SC Nd$_2$CuO$_4$ was found to be intimately linked to those of Ce-doped superconductors as the electrons were doped into the thin films by annealing probably through the creation of oxygen vacancies. Since the electron concentration and superconductivity of the T'-type cuprates are significantly affected by oxygen non-stoichiometry, the electronic structure should be discussed based on the actual electron concentration and oxygen occupancies rather than solely the Ce concentration.

\begin{acknowledgments}
Fruitful discussion with K.~Okada and G.R.~Castro is gratefully acknowledged. Experiments were performed at SPring-8 (proposal Nos.~2015B1699, 2015B1793, 2015B7401, and 2016A1210). This work was supported by Grants-in-aid from the Japan Society of the Promotion of Science (JSPS) (grant Nos.~14J09200 and 15H02109). M.H. acknowledges support from the Advanced Leading Graduate Course for Photon Science (ALPS) and the JSPS Research Fellowship for Young Scientists.
\end{acknowledgments}


\end{document}


\title{Supplementary Information \\ \vspace{1cm} Electronic Structure of Ce-Doped and -Undoped Nd$_2$CuO$_4$ Superconducting Thin Films Studied by Hard X-ray Photoemission and Soft X-ray Absorption Spectroscopy \vspace{1cm}}

\author{M. Horio$^1$, Y. Krockenberger$^2$, K. Yamamoto$^3$, Y. Yokoyama$^3$, K. Takubo$^3$, Y. Hirata$^3$, S. Sakamoto$^1$, K. Koshiishi$^1$, A. Yasui$^4$, E. Ikenaga$^4$, S. Shin$^3$, H. Yamamoto$^2$, H. Wadati$^3$, and A. Fujimori$^1$}

\begingroup
\let\clearpage\relax
\let\vfil\relax
\maketitle
\endgroup

\noindent
{\it \footnotesize $^1$Department of Physics, University of Tokyo, Bunkyo-ku, Tokyo 113-0033, Japan} \\
{\it \footnotesize $^2$NTT Basic Research Laboratories, NTT Corporation, Atsugi, Kanagawa 243-0198, Japan} \\
{\it \footnotesize $^3$Institute for Solid State Physics, University of Tokyo, Kashiwa, Chiba 277-8561, Japan} \\
{\it \footnotesize $^4$Japan Synchrotron Radiation Research Institute, Sayo, Hyogo 679-5198, Japan}

\vfill

\noindent
$^*$e-mail: horio@wyvern.phys.s.u-tokyo.ac.jp \\

\newpage

\section{Conditions for the film growth and annealing (reduction or oxidization)}
All the films discussed here were grown by molecular beam epitaxy (MBE). For typical growth conditions, post-growth annealing under reducing atmospheres is necessary to induce superconductivity in Nd$_{2-x}$Ce$_x$CuO$_4$. The Nd$_{2-x}$Ce$_x$CuO$_4$ ($x=0.15$, 0.19) thin films were grown on (001) SrTiO$_3$ substrates at 680 and 665~$^\circ$C, respectively, using ozone flow rates of 1.0 sccm. After the growth, the films were kept for 10 minutes under ultra-high vacuum (UHV) at 600~$^\circ$C before being subsequently cooled to room temperature. Superconductivity in these Nd$_{2-x}$Ce$_x$CuO$_4$ thin films has been verified also via magnetization measurements, in addition to resistivity measurements [Figs.~1(b) and (c) in the main text]. These superconducting films are labeled as ``SC". The superconducting Nd$_{2-x}$Ce$_x$CuO$_4$ films were divided and one of them was treated under oxidizing conditions using 2.0 sccm ozone at 550~$^\circ$C for 10 minutes and cooled under UHV. These films are labeled as ``oxidized".

The Nd$_2$CuO$_4$ thin films were grown on (001) SrTiO$_3$ substrates at 725~$^\circ$C using ozone flow rate of 1.0 sccm. These films were rapidly cooled down to room temperature. To induce superconductivity in the Nd$_2$CuO$_4$ thin films, {\it ex-situ} annealing in a tubular furnace is necessary. One of the Nd$_2$CuO$_4$ thin films was annealed in a furnace first at 675 $^\circ$C under the O$_2$ partial pressure of $2.5 \times 10^{-4}$ Torr for 1 hour, and then at 525 $^\circ$C under UHV for 10 minutes (annealed Nd$_2$CuO$_4$). Another Nd$_2$CuO$_4$ film was annealed first at 690 $^\circ$C under the O$_2$ partial pressure of $1.0 \times 10^{-3}$ Torr for 1 hour, and then at 500 $^\circ$C under UHV for 10 minutes (weakly annealed Nd$_2$CuO$_4$).

\section{Experimental geometry}
The experimental geometry of hard x-ray photoemission spectroscopy (HAXPES) is described in Supplementary Fig.~\ref{Setup}(a). Measurements were performed with linearly polarized light with grazing incidence to maximize the photoemission intensity \cite{Wadati2013}, and normally emitted photoelectrons were analyzed. The probing depth $\lambda$ is typically 5--20 nm but depends on the kinetic energy of photoelectrons $E_k$ as well as on material \cite{Takata2005}. As for the Cu 2$p$ core level, for example, Panaccione \textit{et al}. \cite{Panaccione2008} derived the equation $\lambda_\mathrm{Cu} = 0.093 E_k^{0.75}$ following experimental estimates for metallic Cu \cite{Sacchi2005} and theoretical calculations using the NIST code \cite{NIST}. This yields $\lambda_\mathrm{Cu} \sim 7$ nm in the present case of $h \nu = 7.94$ eV.

Measurements of soft x-ray absorption spectroscopy (XAS) were carried out in the total electron yield mode using two kind of linearly polarized light, with polar angle $\theta = 90^\circ$  (${\bf E} \perp c$) and $\theta = 30^\circ$, as schematically illustrated in Supplementary Fig.~\ref{Setup}(b). The probing depth $\lambda$ is typically 2--7 nm \cite{Schroeder1996}. As for Cu $L$-edge XAS, Ruosi \textit{et al}. \cite{Ruosi2014} derived $\lambda_\mathrm{Cu} \sim 5$ nm from an incident-photon-angle dependent XAS study on YBa$_2$Cu$_3$O$_7$ thin films.

\begin{figure}
\begin{center}
\includegraphics[width=130mm]{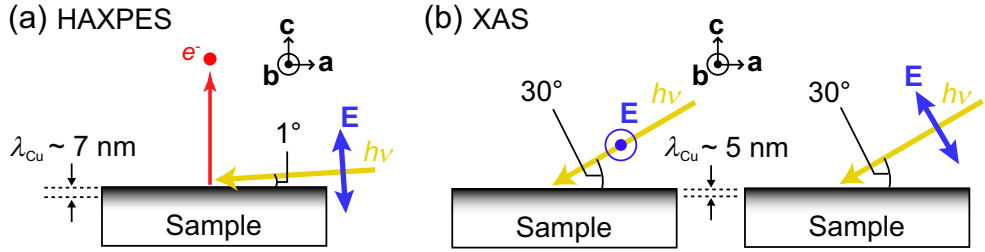}
\end{center}
\caption{Experimental geometries. (a) HAXPES measurements. (b) XAS measurements. Linearly polarized light with polar angle $\theta = 90^\circ$  (${\bf E} \perp c$, left) and $\theta = 30^\circ$ (right) were used.}
\label{Setup}
\end{figure}

\section{Matrix elements of XAS}
Using Fermi's golden rule, matrix elements for Cu $L$-edge absorption relevant to the present study with light polarized along ${\bf n} = (\mathrm{sin}\theta\mathrm{cos}\phi, \mathrm{sin}\theta\mathrm{sin}\phi, \mathrm{cos}\theta)$ can be written as follows \cite{Pellegrin1993}: 

\begin{eqnarray}
\left< 2p_{\mathrm{in-plane}} \left| H' \right| 3d_{x^2-y^2} \right>^2 &=& \sum_{i = x,y} \left< 2p_i \left| H' \right| 3d_{x^2-y^2} \right>^2 = M_{if}^2\mathrm{sin}^2\theta \\
\left< 2p_{\mathrm{in-plane}} \left| H' \right| 3d_{3z^2-r^2} \right>^2 &=& \sum_{i = x,y} \left< 2p_i \left| H' \right| 3d_{3z^2-r^2} \right>^2 = \frac{1}{3}M_{if}^2\mathrm{sin}^2\theta \\
\left< 2p_z \left| H' \right| 3d_{x^2-y^2} \right>^2 &=& 0 \\
\left< 2p_z \left| H' \right| 3d_{3z^2-r^2} \right>^2 &=& \frac{2}{3}M_{if}^2\mathrm{cos}^2\theta,
\end{eqnarray}
where $H'$ is the Hamiltonian for electron-photon interaction, and $M_{if}$ is a reduced matrix element. Therefore, absorption with polarization perpendicular to the $c$ axis (${\bf E} \perp c$, $\theta = 90^\circ$) is 75 \% due to the transition into the 3$d_{x^2-y^2}$ orbital, and that with polarization parallel to the $c$ axis (${\bf E} \parallel c$, $\theta = 0^\circ$) is entirely due to transitions into the 3$d_{3z^2-r^2}$ orbital.

\section{Calculations of O 1\mbox{\boldmath $s$} binding energies}
The O $1s$ binding energies observable by photoemission spectroscopy were estimated from the calculation of Slater transition states \cite{Slater1971} using density functional theory (DFT). Calculations were carried out using the generalized gradient approximations (GGA). For the 4$f$ orbital, since GGA generally yields unrealistically high density of states (DOS) in a narrow energy window around $E_\mathrm{F}$, which could affect the O $1s$ binding energy, we have carried out three different kinds of calculations: \\
(i): Nd$_2$CuO$_4$ with GGA \\
(ii): La$_2$CuO$_4$ with the lattice constant of Nd$_2$CuO$_4$ with GGA \\
(iii): Nd$_2$CuO$_4$ with GGA+$U$ ($U_{\mathrm{4}f} = 9$ eV)\\
in (ii), Nd is replaced by La, where 4$f$ levels are unoccupied, and the 4$f$ level is located far above $E_\mathrm{F}$. In (iii), on-site Coulomb repulsion $U = 9$ eV is introduced to the Nd 4$f$ orbitals thereby splitting the 4$f$ DOS away from $E_\mathrm{F}$. All the calculations were performed with spin polarization and the O $1s$ binding energies were finally determined by taking the average of spin-down and -up binding energies. The calculated O $1s$ binding energies are listed in Table.~\ref{O1s}. Although the magnitude of the difference between the O$_\mathrm{Nd_2O_2}$ and O$_\mathrm{CuO_\mathrm{2}}$ 1$s$ binding energies depends on the calculation method, O$_\mathrm{Nd_2O_2}$ always resides at a lower binding energy and O$_\mathrm{CuO_\mathrm{2}}$ at a higher binding energy.

\begin{table}[htb]
	\begin{center}
	{\tabcolsep = 4mm
	\begin{tabular}{ccc}
		\hline \hline
			Calculation method & O$_\mathrm{Nd_2O_2}$ (eV) & O$_\mathrm{CuO_\mathrm{2}}$ (eV)  \\  \hline 
		(i) & 527.93  & 528.36  \\
		(ii) & 527.81  &  528.38  \\
		(iii) & 527.54  &  528.53  \\ \hline \hline
	\end{tabular}}
	\end{center}
	\caption{O $1s$ binding energies estimated from DFT calculations. Details of the calculations are described in the text.}
	\label{O1s}
\end{table}

\section{Procedure for fitting O 1\mbox{\boldmath $s$} peaks}
First, the O 1$s$ spectrum of as-grown Nd$_2$CuO$_4$ was fitted to a superposition of three peaks: Voigt function (for O$_\mathrm{Nd_2O_2}$), Mahan line shape $\frac{1}{\mathrm{\Gamma}(\alpha)} \frac{e^{-(E_B-E_0)/\xi}}{\left| (E_B-E_0)/\xi \right| ^{1-\alpha}} \Theta(E_B-E_0)$ convolved with a Voigt function (for O$_\mathrm{CuO_\mathrm{2}}$), and Another Voigt function (for contamination peak at $\sim 531$ eV). The full width at half maximum (FWHM) of the Lorentian and Gaussian in the Voigt function was assumed to be the same for O$_\mathrm{Nd_2O_2}$ and O$_\mathrm{CuO_\mathrm{2}}$ peaks. The parameter $\xi$ in the Mahan line shape, which should be a magnitude of the order of the Fermi energy, was fixed at 1 eV, which was determined from nearest-neighbor hopping parameter $t \sim 0.25$ eV previously derived for Nd$_{2-x}$Ce$_x$CuO$_4$ ($x = 0.15$) \cite{Ikeda2009}. The fitting yielded O$_\mathrm{Nd_2O_2}$ and O$_\mathrm{CuO_\mathrm{2}}$ peaks with the area ratio of 0.44 : 0.56, consistent with the fact that the number of oxygen atoms in the Nd$_2$O$_2$ layers and the CuO$_2$ planes are the same. The obtained FWHM of the Lorentzian was 0.15 eV, which is close to the inherent O $1s$ core-hole lifetime of 0.16 eV observed for H$_2$O \cite{Sankari2003}, while the FWHM of the Gaussian was 0.75 eV, which is larger than the experimental total resolution of 0.3 eV possibly due to the contribution from phonons \cite{Citrin1974}.

Then, the O $1s$ spectrum of annealed Nd$_2$CuO$_4$ was fitted to the same functional form with the FWHM of the Lorentzian and the Gaussian fixed to the values of the as-grown film. The parameter $\xi$ was again fixed at 1 eV. The area ratio between O$_\mathrm{Nd_2O_2}$ and O$_\mathrm{CuO_\mathrm{2}}$ peaks obtained by fitting was 0.46 : 0.54, which is close to that for the as-grown film, and hence validating the result of the fitting.

\begin{figure}
\begin{center}
\includegraphics[width=165mm]{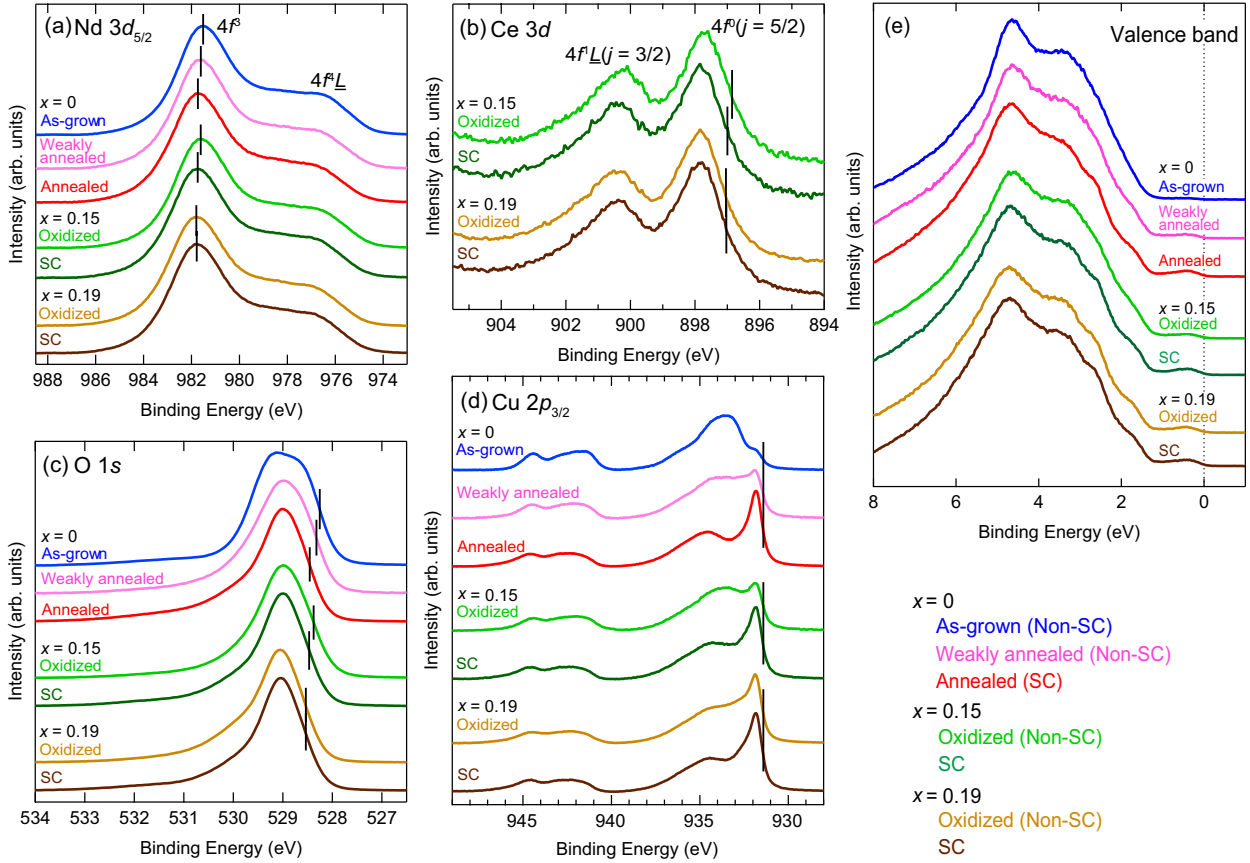}
\end{center}
\caption{HAXPES spectra of Nd$_{2-x}$Ce$_x$CuO$_4$ thin films. (a) Nd 3$d_\mathrm{5/2}$. (b) Ce 3$d$. (c) O 1$s$. (d) Cu 2$p_{3/2}$. (e) Valence band. Peak or edge positions used to determine the core-level shifts are indicated by vertical bars.}
\label{Nd_O_Ce_Cu}
\end{figure}

\section{HAXPES spectra of Nd$_{2-x}$Ce$_x$CuO$_4$ films}
In Supplementary Fig.~\ref{Nd_O_Ce_Cu}, core-level HAXPES spectra for Nd$_2$CuO$_4$ (as-grown, weakly annealed, annealed) and Nd$_{2-x}$Ce$_x$CuO$_4$ ($x = 0.15, 0.19$, SC, oxidized) are plotted. The Nd and Ce peak positions are shifted between films, but their spectral line shapes are almost identical. On the other hand, the O 1$s$ peaks change their shape with varying Ce concentration and oxygen content. Because the changes in the O 1$s$ peaks can be understood within the above scenario, we compare the low energy edges of different films as representing the O$_\mathrm{Nd_2O_2}$ peak position, and summarize them in Supplementary Fig,~\ref{muShift}(a). Not only the Nd 3$d$ and O 1$s$ core-level peaks, but also the Ce 3$d$ peaks are shifted by the same amount for every film (The origins of the shifts of the Ce 3$d$ were set at the Nd 3$d$ shift for the $x = 0.15$ oxidized film.), confirming that those core-level shifts corresponds to the chemical-potential shift. 

While the Nd, O, and Ce core levels were shifted by a few hundreds meV, the shifts observed for the edge of the lowest energy peak in the Cu 2$p$ core-level spectrum was rather small as plotted in Supplementary Fig.~\ref{muShift}. Considering that the lowest energy peak corresponds to the final state where a core hole is screened by conduction electrons, the behavior can be understood within a simple Kotani-Toyozawa picture \cite{Kotani1974}. When the chemical potential is shifted, core-hole energy increases accordingly, but energy gain by transferring an electron from conduction band to Cu 3$d$ level at the core-hole site also increases by the same amount, compensating the chemical-potential shift. The binding energy of the Cu 2$p$ lowest energy edge is thus insensitive to the amount of carriers.

\begin{figure}
\begin{center}
\includegraphics[width=105mm]{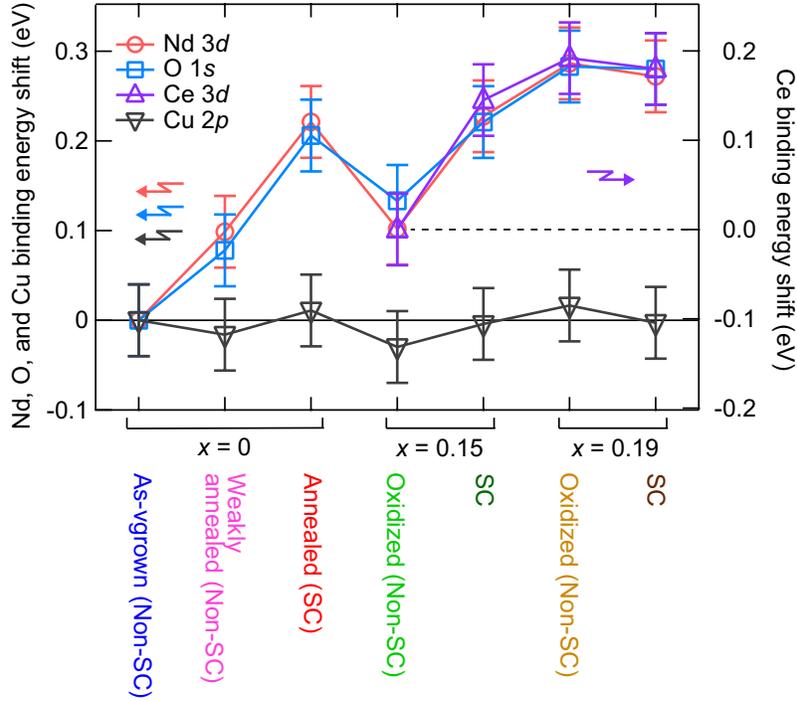}
\end{center}
\caption{Binding energy shifts of the Nd 3$d$, O 1$s$, Ce 3$d$, and Cu 2$p$ core levels. The shifts were estimated by the peak position for Nd 3$d$ and by the low binding-energy edge for O 1$s$, Ce 3$d$, and Cu 2$p$. The origin of the shifts of Ce 3$d$ are set at the shift of Nd 3$d$ for the $x = 0.15$ oxidized film.}
\label{muShift}
\end{figure}
